\documentclass{jpp}
\usepackage{graphicx}
\usepackage{natbib}
\usepackage[T1]{fontenc}
\usepackage{bm}
\usepackage{color}
\usepackage{amsmath}
\usepackage{amssymb}
\usepackage{amsfonts}
\usepackage{courier}
\usepackage{txfonts}

\newcommand{\appref}[1]{\hyperref[#1]{Appendix~\ref{#1}}}

\usepackage{ae}
\usepackage{units}
\usepackage[disable]{todonotes}

\newcommand{\be}{\begin{displaymath}}
\newcommand{\ee}{\end{displaymath}}
\newcommand{\bn}{\begin{equation}}
\newcommand{\en}{\end{equation}}

\usepackage{enumerate}

\usepackage[T1]{fontenc}
\usepackage{bm}
\usepackage{color}
\usepackage{graphicx}
\usepackage{multirow}
\usepackage{multicol}
\usepackage[breaklinks=true]{hyperref}
\hypersetup{
  unicode=false,          
  pdftoolbar=true,        
  pdfmenubar=true,        
  pdffitwindow=false,     
  pdfstartview={FitH},    
  pdfsubject={Subject},   
  pdfcreator={Creator},   
  pdfproducer={Producer}, 
  pdfkeywords={keyword1} {key2} {key3}, 
  pdfnewwindow=true,      
  colorlinks=true,        
  linkcolor=blue,         
  citecolor=blue,         
  filecolor=blue,         
  urlcolor=blue           
}

\usepackage{amsmath}
\usepackage{gensymb}
\usepackage[normalem]{ulem}

\title{Effects of oblique incidence and colliding pulses on laser-driven proton acceleration from relativistically transparent ultrathin targets}

\author{J.~Ferri\aff{1}\corresp{\email{julienf@chalmers.se}}, E.~Siminos\aff{2}, L.~Gremillet\aff{3} \and T.~F\"ul\"op\aff{1}} 

\affiliation{\aff{1}Department of Physics, Chalmers University of Technology,
 SE-41296 G\"{o}teborg, Sweden
   \aff{2}Department of Physics, University of Gothenburg, SE-41296 G\"{o}teborg, Sweden  \aff{3}CEA, DAM, DIF, F-91297 Arpajon, France}

\begin{document}

\maketitle

\begin{abstract}
  The use of ultrathin solid foils offers optimal conditions for accelerating protons from laser-matter interactions. When the target is thin enough that relativistic self-induced transparency (RSIT) sets in, all of the target electrons get heated to high energies by
  the laser, which maximizes the accelerating electric field and therefore the final ion energy. In this work, we first investigate how ion acceleration by ultraintense femtosecond laser pulses in transparent CH$_2$ solid foils is modified when turning from normal to oblique ($45^\circ$) incidence. Due to stronger electron heating, we find that higher proton energies can be obtained at oblique incidence but in thinner optimum targets.  
  We then show that proton acceleration can be further improved by splitting the laser pulse into two half-pulses focused at opposite incidence angles. An increase by $\sim 30\,\%$ in the maximum proton energy and by a factor of $\sim 4$ in the high-energy proton charge is reported compared to the reference case of a single normally incident pulse.
 \end{abstract}
\section{Introduction}
\indent

Laser-driven proton acceleration using ultraintense pulses has been an active field of research in the past two decades~\citep{Tikhonchuk2010, Daido2012, Macchi2013}. The strong electric fields produced in the interaction of the ultraintense laser pulse with solid targets can accelerate protons to high energies on acceleration distances of only a few microns. This has led to the development on small-scale facilities of a large variety of applications, such as radiography~\citep{Romagnani2005}, generation on warm dense matter~\citep{Patel2003,Pelka2010}, inertial confinement fusion~\citep{roth2001}, nuclear physics~\citep{McKenna2003}, or proton therapy \citep{Bulanov2002}. The most investigated, and routinely exploited mechanism is the so-called target normal sheath acceleration (TNSA), which benefits from a relatively easy implementation~\citep{Wilks2001, Clark2000, Maksimchuk2000, Snavely2000}. TNSA, however, yields maximum ion energies that scale rather weakly with the laser intensity ($E_{\rm max} \propto \sqrt{I_0}$) \citep{Daido2012}. This limitation motivates the investigation of alternative, possibly more efficient schemes, \emph{e.g.}, radiation pressure (or light-sail) acceleration~\citep{Esirkepov2004, Kar2012}, shock acceleration~\citep{Silva2004, Haberberger2012}, directed Coulomb explosion~\citep{Bulanov2005} or breakout afterburner~\citep{Yin2006}. 

The possibility of employing several laser pulses to control the preplasma formation and increase the general efficiency of the TNSA process has been studied in experiments~\citep{Markey2010, Scott2012, Brenner2014, Ferri2018}. In particular, in a recent work, we studied the influence of splitting the laser pulse in two pulses containing half of the total energy in a TNSA setup \citep{Ferri2019}. In that work, the half-pulses were simultaneously focused onto a few-$\mu$m thick target, but at opposite incidence angles. Through particle-in-cell (PIC) simulations, we showed that the proton energy could be increased in the two-pulse scheme by about 80\% with lasers impinging at $\pm 45^\circ$ angles onto a sharp-gradient target having sub-wavelength preplasma scale-length. Such enhancement was found to result from a more efficient Brunel-type vacuum heating of the target electrons, due to the laser interference and the suppression of the mitigating $v\times B$ force at the target surface \citep{Geindre2006}.

While those results are very promising for TNSA experiments, notably those performed at ultrahigh laser contrast~\citep{Ceccotti2007}, they also suggest possible improvements for other acceleration setups, especially those involving ultrathin (nanometric) target foils. Previous studies indeed demonstrated that, in the case of femtosecond laser pulses, ion acceleration is optimized for target thicknesses such that relativistic self-induced transparency (RSIT) occurs near the laser peak~\citep{Esirkepov2006, Dhumieres2013, Brantov2015}. It is therefore worth exploring whether the use of two interfering half-pulses in this regime could significantly enhance the hot-electron generation, and hence lower the RSIT threshold.

In this work, we investigate by means of PIC numerical simulations,
how splitting the laser pulse into two half-pulses focused at opposite incidence angles affects the interaction. For the purpose of establishing a reference case,
we first investigate the dynamics of nanometric $\rm CH_2$ foils driven by an ultraintense laser pulse under normal or oblique incidence. While the latter case leads to higher proton energies due to stronger electron heating, the optimum foil target is about twice thinner than that found at normal incidence, a consequence of self-induced magnetostatic fields within the ion acceleration region.
Secondly, we find that, by further enhancing the electron heating and mitigating the growth of the deleterious magnetostatic fields, the two-pulse scheme significantly increases the cutoff energy and number of the accelerated protons. It may also facilitate experimental investigations of RSIT-related processes by allowing thicker targets to turn relativistically transparent.

\section{Proton acceleration at normal or oblique laser incidence in relativistically transparent foils}
\indent

\begin{figure}
\centering
\includegraphics[width=1\textwidth]{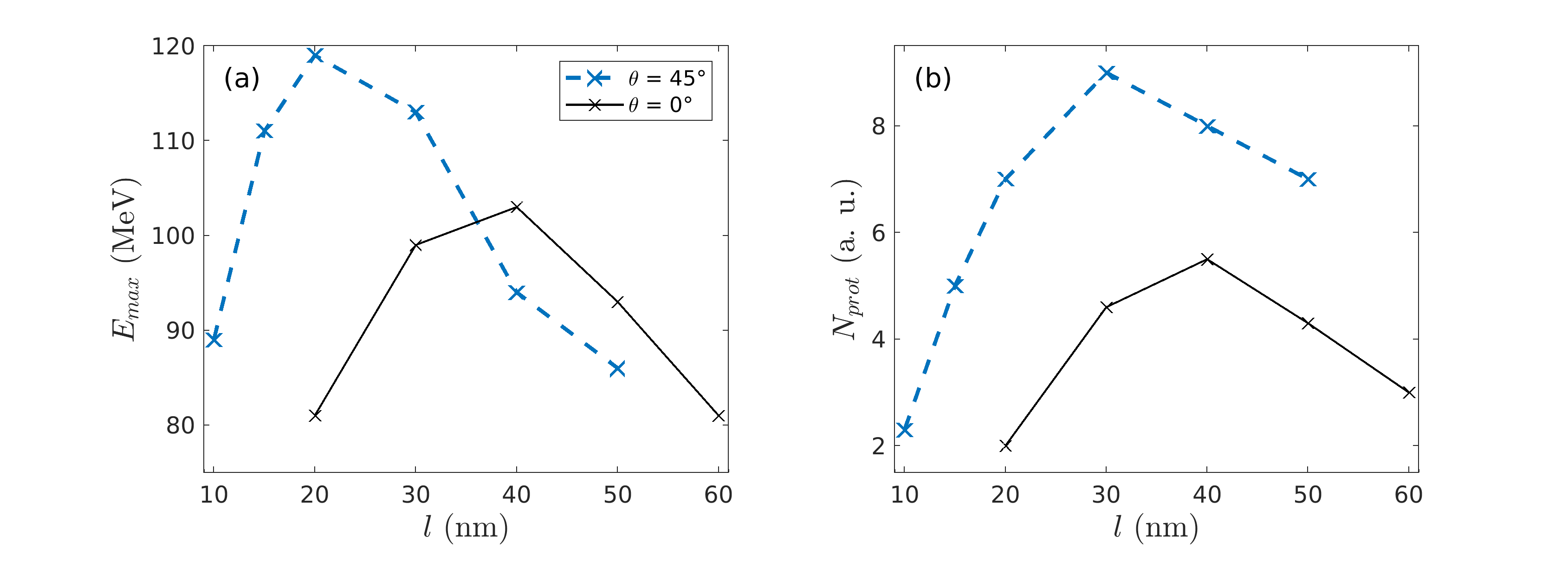}
\caption{Ion acceleration by a single, $16\,\rm J$ energy, $38\,\rm fs$ duration laser pulse as a function of the CH$_2$ foil thickness: (a) maximum proton energy and (b) number of protons with $>30\,\rm MeV$ energies. The laser pulse is either normally incident (solid black) or obliquely incident at a $45^\circ$ angle (dashed blue).
}
\label{prot1p}
\end{figure}

According to  previous works \citep{Esirkepov2006, Dhumieres2013, Brantov2015}, optimal ion acceleration by femtosecond laser pulses is achieved for target thicknesses close to the RSIT threshold, $l_{\rm opt} \simeq \lambda_0 a_0 n_c/2n_e$, where $\lambda_0$ and $a_0$ are, respectively, the laser wavelength and maximum potential vector, $n_c$ is the critical density and $n_e$ is the electron density of the target.

Our simulations have been performed in 2D geometry using the PIC code \textsc{epoch} \citep{Arber2015}.  As a reference configuration, we have first considered a single laser pulse of $16\,\rm J$ energy and $0.8\,\rm \mu$m wavelength interacting at normal incidence with a solid-density CH$_2$ foil. The pulse is $p$-polarized (\emph{i.e.}, along the $y$ axis), has Gaussian transverse and temporal profiles of FWHM $w_0=7\,\rm \mu m$ and $\tau_0 =38\,\rm fs$, respectively, and a peak intensity $I_0 = 5\times 10^{20}\,\rm Wcm^{-2}$ ($a_0 \simeq 15$).
The reference time $t=0$ corresponds to the pulse maximum reaching the target. The CH$_2$ target is modelled as a fully ionized plasma at a $1.1\,\rm gcm^{-3}$ density, corresponding to a total electron density $n_e = 220n_c$. Its thickness, $l$, is varied in a range, $20 \le l \le 60\,\rm nm$, encompassing the predicted optimum, $l_{\rm opt} = 30\,\rm nm$. We use 200 particles per cell and per species. The simulation domain has dimensions of $75\times 80\,\rm \mu m^2$ with mesh size $\Delta x \times \Delta y = 4 \times 10\,\rm nm^2$.

Figures~\ref{prot1p}(a,b) plot (as black solid curves) the target thickness dependence of the (a) maximum proton energy and (b) the number of high-energy (above 30~MeV) protons, as measured at the final simulation time ($t_{\rm max}=300\,\rm fs$). Both quantities are found to be maximized at $l=40\,\rm nm$, close to $l_{\rm opt}$, with proton energies as high as $\sim 100\,\rm MeV$ being then recorded. The number of high-energy protons is more sensitive to the target thickness than their maximum energy. Using $l=20\,\rm nm$ or $60\,\rm nm$ leads to a decrease in the proton number by a factor of 2 to 3 compared to the optimum case. 

\begin{figure}
\centering
\includegraphics[width=1\textwidth]{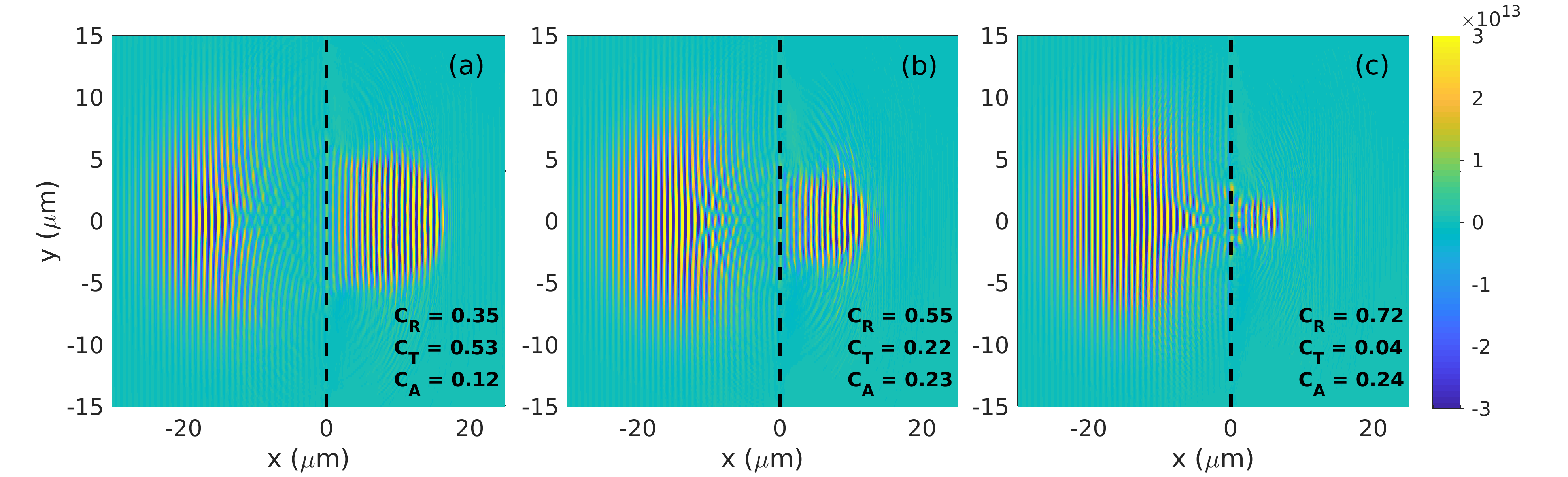}
\caption{Transverse electric field $E_y$ (in $\rm Vm^{-1}$) at $t=45\,\rm fs$ after the on-target laser peak, for foil thicknesses (a) $l=20\,\rm nm$, (b) $l=40\,\rm nm$ (b) and (c) $60\,\rm nm$. In each panel the transmitted ($C_T$), reflected ($C_R$) and absorbed ($C_A$) fractions of the laser energy are noted, and the dashed black line indicates the initial position of the foil.}
\label{Eyfield}
\end{figure}

The optimum acceleration conditions are reached when RSIT occurs near the laser peak, which is confirmed by looking at the laser fields going through the target in Figs.~\ref{Eyfield}(a-c), and by computing, for each thickness value, the transmitted ($C_T$), reflected ($C_T$) and absorbed ($C_A$) fractions of the laser energy. While the laser pulse is significantly transmitted ($C_T \simeq 0.53$) and relatively weakly absorbed ($C_A \simeq 0.12$) through the $20\,\rm nm$ foil [Fig.~\ref{Eyfield}(a)], it gets significantly reflected ($C_R \simeq 0.55$) when the foil thickness is increased to $40\,\rm nm$ [Fig.~\ref{Eyfield}(b)], while its absorption is approximately doubled ($C_A\simeq 0.23$). If the foil thickness is further increased to $60\,\rm nm$, the laser absorption remains similar, $C_A \simeq 0.24$, but the transmission drops to a very low value, $C_T \simeq 0.04$ [Fig.~\ref{Eyfield}(c)]. As a result of the RSIT then taking place late in the decreasing part of the pulse, the expanding protons no longer benefit from the boost in energy caused by volumetric electron heating at the laser peak \citep{Mishra2018}, and hence experience a weaker overall acceleration. This corresponds to the transition to the TNSA regime.

The absorption processes, and therefore the onset of RSIT, are expected to be altered when operating at oblique laser incidence. To assess the dependency of ion acceleration on the incidence angle, we now consider a laser pulse impinging at an angle $\theta = 45^\circ$ onto the foil, all other physical parameters being kept constant. The simulation domain is also enlarged to $75 \times 120~\rm \mu m^2$. Figures~\ref{prot1p}(a,b) show (as blue dashed curves) that the optimum target thickness in terms of the maximum proton energy (resp. proton number above $30\,\rm MeV$) then drops from $l=40\,\rm nm$ to $20\,\rm nm$ (resp. $30\,\rm nm$). This decrease seems reasonable since an oblique incidence angle leads to an increased interaction length ($l_{\rm eff} = l/\cos \theta$). 
More surprisingly, the decrease in the optimum target thickness goes along with a significant enhancement of the proton cutoff energy (by $\sim 15\%$, up to $\sim 120\,\rm MeV$) and number (by $\sim 60\%$).

\begin{figure}
\centering
\includegraphics[width=1\textwidth]{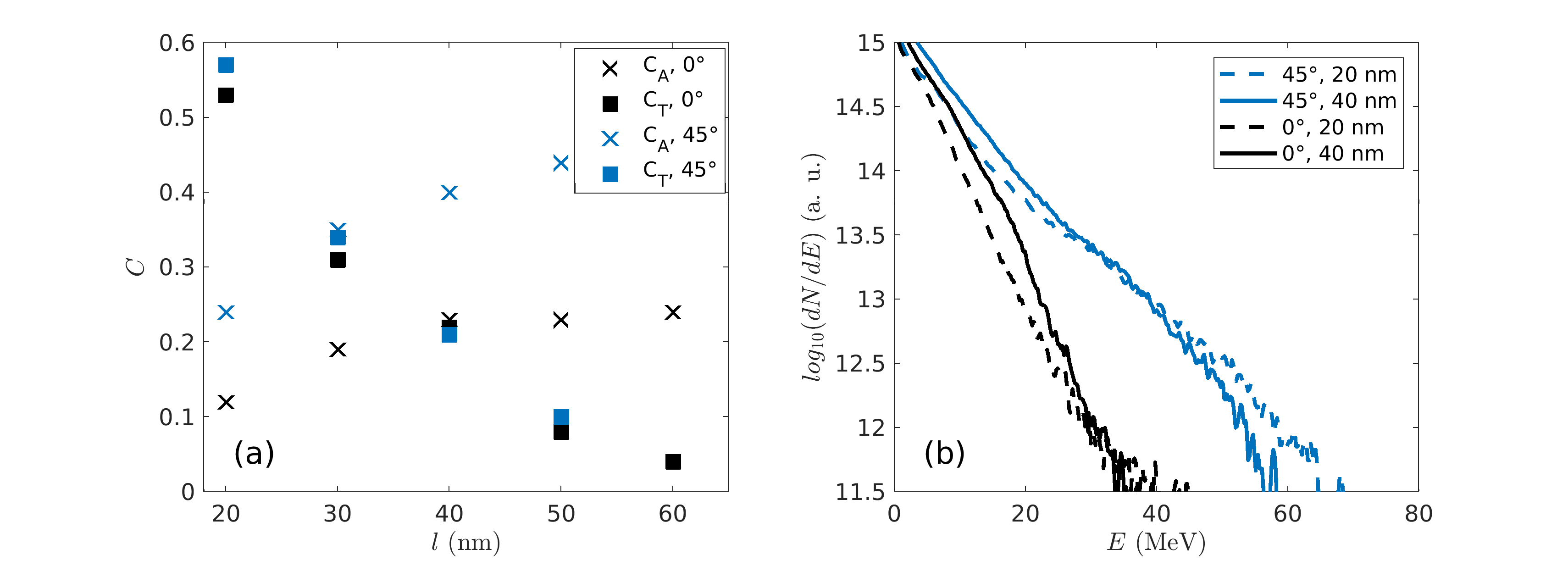}
\caption{(a) Absorption ($C_A$, crosses) and transmission ($C_T$, squares) coefficients for normal (black) and oblique (blue) incidence as a function of the target thickness. (b) Electron energy distribution recorded at $t=20\,\rm fs$ after the on-target laser peak; black and blue curves correspond to normal and oblique ($45^\circ$) laser incidence, while solid and dashed curves correspond to $40\,\rm nm$ and $20\,\rm nm$ thick targets.
}
\label{abs1p}
\end{figure}

The improvement of the proton acceleration at oblique incidence can be explained by a more efficient mechanism for hot-electron generation. The absorption and transmission coefficients for the normal and oblique incidence are summarised in Fig.~\ref{abs1p}(a), showing an overall (approximately twofold) increase in the absorption coefficient at oblique incidence. For instance, the absorption coefficient at $l=40\,\rm nm$ rises from $C_A \simeq 0.23$ at $\theta=0^\circ$ to $C_A \simeq 0.40$ at $\theta = 45^\circ$.
By contrast, the transmission coefficient vs. thickness values are similar at $\theta = 0^\circ$ and $45^\circ$ indicating that the transparency condition weakly depends on the incidence angle.

Figure~\ref{abs1p}(b) plots the electron energy spectra recorded $20\,\rm fs$ after the on-target pulse maximum for two values of the the target thickness ($l=20\,\rm nm$ and $40\,\rm nm$).
They show that the higher laser absorption found at oblique incidence translates into an increase in both the number and temperature of the hot ($>1\,\rm MeV$) electrons. The thicker (40~nm) target seems to mainly boost the number of hot electrons in both cases. 

To understand the simulation results we adapt the thin-foil expansion model proposed in \citep{Brantov2015} to the considered scenario. This model considers the expansion of a neutral plasma slab of initial thickness $l$ and uniform ion density $n_{i0}$. It assumes a Boltzmann electron distribution, $n_e(x,t) = n_e(0,t) \exp [e \phi(x,t)/T_e(t)]$, where $\phi(x,t)$ is the electrostatic potential, $T_e(t)$ is the (longitudinal) electron temperature and $n_e(0,t)$ is the (time-decreasing) electron density at the target center ($x=0$). The electrostatic field $E_x = -\partial_x \phi$ is taken to vary linearly inside the ion-filled region: $E_x(x,t) = E_f x/x_f(t)$ for $0 \le x \le x_f(t)$, where $x_f(t)$ is the position of the ion front [$x_f(0)=l/2$] and $E_f(t)$ is the field strength at the ion front. From this it readily follows that $\phi(x,t) = -E_f(t) x^2/2x_f(t)$. Outside the ion cloud ($x>x_f$), the electric potential obeys the nonlinear Poisson equation, $\partial_{xx} \phi = en_e(0,t) \exp(e\phi/T_e)/\epsilon_0$. Multiplying this equation by $\partial_x \phi$ and integrating it over $[x_f,\infty]$ yields $n_e(0,t) = ( \epsilon_0 E_f^2/2T_e) \exp(eE_f x_f/2 T_e)$. Furthermore, integrating Gauss's equation over $[0,x_f]$ gives
\begin{align}
    E_f(t) &= \frac{Z e n_{i0}l}{2\epsilon_0} - \frac{e n_e(0,t)}{\epsilon_0} \int_0^{x_f} dx\, e^{e \phi/T_e} \nonumber \\
&= \frac{Z e n_{i0}l}{2\epsilon_0} - \frac{e E_f^2}{2T_e} e^{eE_f x_f/2T} \int_0^{x_f} dx\,e^{-eE_fx^2/2x_f T_e} \,,
\end{align}
where $Z$ is the ion charge state. Introducing $\mathrm{erf}(z)$ the error function, the above equation can be recast as
\begin{equation} \label{eq:E_f}
    E_f(t) = \frac{Zen_{i0}l}{2 \epsilon_0} - \frac{\sqrt{\pi} E_f}{2} \sqrt{\frac{e E_f x_f }{2 T_e}} \mathrm{erf} \left( \sqrt{\frac{e E_f x_f }{2 T_e}}\right) \,.
\end{equation}
This formula assumes a 1D expansion geometry, and hence should overestimate the field strength when the ion front has moved a distance comparable with the transverse size of the sheath field ($\sim w_0 \cos \theta$). A correction factor is thus applied to account for this effect:
\begin{equation} \label{eq:E_f_mod}
    E_{f,\delta}(t) = \frac{E_f(t)}{\left[ 1+ \left(\frac{2 x_f(t)}{w_0 \cos \theta} \right)^2\right]^{(\delta -1)/2}},
\end{equation}
where $\delta \in (2,3)$ is the spatial dimensionality of the problem.

The electron temperature $T_e(t)$ is assumed to follow the rise in the laser intensity $I_0(t)$, here taken in the form $\sin^2(\pi t/2\tau_0)$, and to stay constant in its decreasing part.
The maximum electron temperature is inferred from the measured laser absorption: $T_{e,\rm max}/m_ec^2 = 0.8C_A (a_0^2/2) (c \tau_0 /l \cos \theta) (n_c/n_e)$, where the $0.8$ factor corresponds to an averaging of the intensity over the laser focal spot. After the irradiation, the electrons start cooling down adiabatically. More specifically, $T_e(t)$ evolves as
\begin{equation}
    T_e(t)/T_{e,\rm max} =
    \left\{
    \begin{array}{ll}
      \sin^2\left(\pi t/2\tau_0\right) & t \le \tau_0 \, \\
      1 & \tau_0 < t \le 2\tau_0 \,,\\
      \left[1+\left(\frac{c_s(t-2\tau_0)}{x_f(2\tau_0)} \right)^{\Gamma-1}\right]^{-1} & t > 2\tau_0 \,.
    \end{array} \right.
\end{equation}
We have introduced $\Gamma (T_e)$ the generalized adiabatic index for a relativistic 1D gas and the typical sound speed $c_s = \sqrt{ZT_{e,\rm max}/m_i}$.

The knowledge of $E_f(t)$ allows the velocity and position of the front ions to be advanced through
\begin{align} \label{eq:motion}
    &\frac{dv_f(t)}{dt} = Ze E_f(t)/m_i \,\\
    &\frac{dx_f(t)}{dt} = v_f(t) \,.
\end{align}
It is to be noted  that the model assumes a single ion species of charge state $Z$ and mass $m_i = A m_p$ ($m_p$ is the proton mass). Although the expansion dynamics of light ions may be affected by that of heavier ions in a composite target~\citep{Brantov2006}, we will take $Z=1$ and $A=1$ in the following.

\begin{table}
    \centering
    \begin{tabular}{*{5}{c}}
         $l$ (nm) &20 & 30 & 40 & 50\\ 
         $E_{\rm max,0\degree}$ (MeV) & $100$  & $119$  & $119$  & $111$ \\ 
         $E_{\rm max,45\degree}$ (MeV) & $139$ & $156$ & $156$ & $151$\\ 
    \end{tabular}
    \caption{Maximum proton energy produced by an adaptation of the model from \cite{Brantov2015} for varying target thicknesses, at normal ($E_{\rm max,0\degree}$) and oblique ($E_{\rm max,45\degree}$) incidence.}
    \label{tab:tab1}
\end{table}

Equations~\eqref{eq:E_f}-\eqref{eq:motion} have been solved numerically for $20 \le l \le 50\,\rm nm$ and $\theta \in (0^\circ,45^\circ)$.
The results are gathered in Table~\ref{tab:tab1}. The values obtained at normal incidence are quite close to those observed in the simulations -- although generally slightly overestimated -- and reproduce the same trend, with an optimal target thickness between 30 and 40~nm. At oblique incidence, however, the model consistently overestimates the maximum proton energies, especially in the thicker targets ($l\ge 40\,\rm nm$). It also predicts an optimal thickness of $30-40\,\rm nm$, larger than the one ($\sim 20\,\rm nm$) observed in the simulations. This discrepancy between the simulations and the model suggests that the latter lacks a saturation mechanism that is mainly operative at oblique incidence and in thicker targets.

\begin{figure}
\centering
\includegraphics[width=1\textwidth]{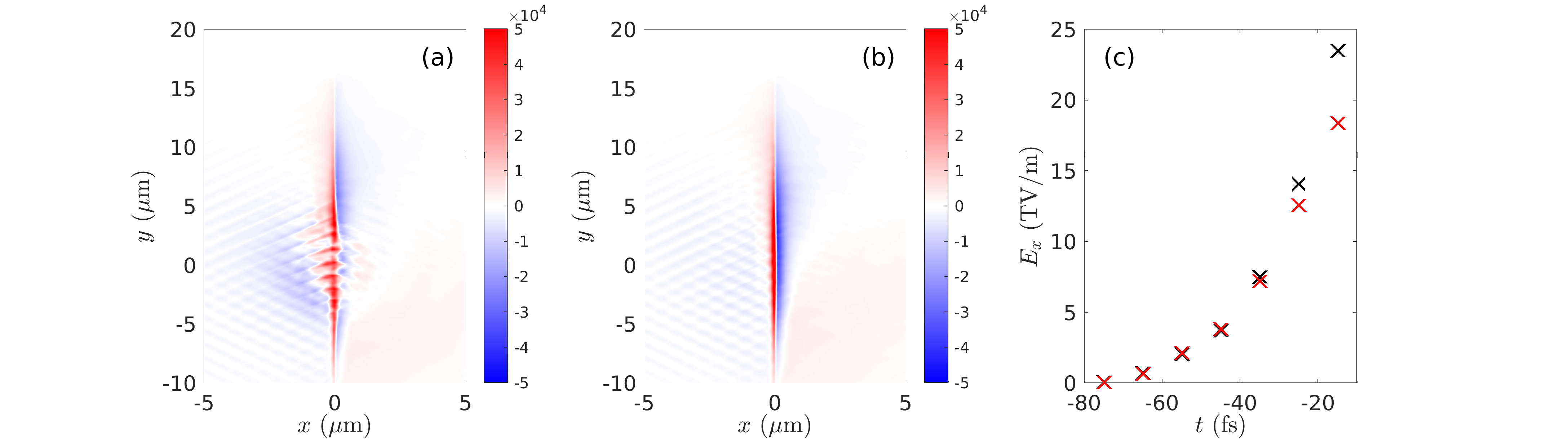}
\caption{(a,b) Magnetostatic $B_z$ field (in T) $10\,\rm fs$ before the on-target laser peak for a single obliquely incident pulse and target thicknesses (a) $l=20\,\rm nm$ and (b) $l=40\,\rm nm$. (c) Maximum of the ion-accelerating, electrostatic field $E_x$ on axis at the rear surface before the on-target peak, for foil thicknesses 20~nm (black) and 40~nm (red).}
\label{Bz220}
\end{figure}

One such saturation-inducing effect is the generation of a strong magnetostatic field $B_z$ along the target surfaces [see Figs.~\ref{Bz220}(a,b), associated with $l=20\,\rm nm$ and 40~nm], due to the electron current driven in the target in the transverse ($y$) direction. When volumetric electron acceleration is achieved (as is the case in the RSIT regime), this field is expected to increase with the target thickness [compare Figs.~\ref{Bz220}(a) and \ref{Bz220}(b), where the $B_z$ field strength is seen to rise from $\sim 2\times 10^4\,\rm T$ at $l=20\,\rm nm$ to $\sim 4\times 10^4\,\rm T$ at $l=40\,\rm nm$ around the laser spot]. This field tends to deflect the hot electrons in the transverse direction, hence limiting the growth of the on-axis electrostatic sheath field ($E_x$) already before the on-target pulse maximum. This is evidenced in Fig.~\ref{Bz220}(c), which plots, for both $l=20\,\rm nm$ and 40~nm, the temporal evolution of the maximum value of $E_x$ on axis at the rear surface. A similar magnetic effect has previously been shown experimentally to inhibit TNSA in micrometric targets~\citep{Nakatsutsumi2018}. Therefore, although they are expected to produce stronger sheath fields, the thicker targets that turn transparent at $\theta=45^\circ$ do not yield faster ions, ultimately resulting in an optimum target thickness that is about half that found at $\theta=0^\circ$.

\section{Proton acceleration with colliding laser pulses in relativistically transparent foils.}
\indent
We now turn to a configuration where
the main laser pulse (of total energy of 16~J) is split into two half-pulses of equal energy (8~J) incident on the target at opposite angles ($\pm 45^\circ$), similarly to the lower-intensity setup that was lately found to boost TNSA \citep{Ferri2019}.
The other parameters remain the same as in the previous section.

\begin{figure}
\centering
\includegraphics[width=1\textwidth]{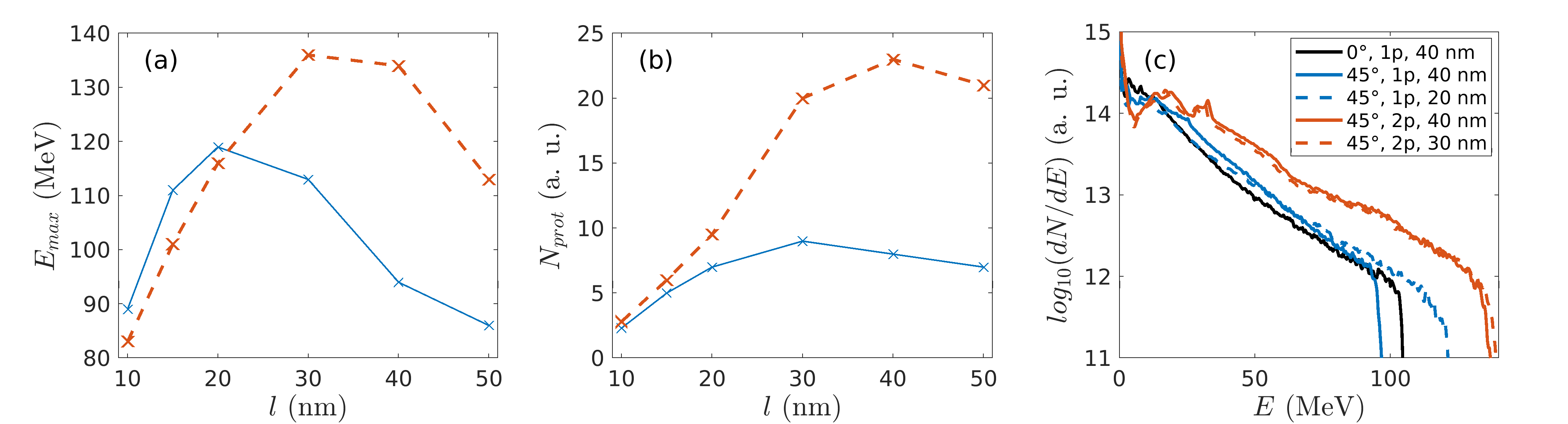}
\caption{(a,b) Ion acceleration by two $8\,\rm J$, $38\,\rm fs$ half-pulses incident at $\pm 45^\circ$ angles (dashed red) as a function of the CH$_2$ foil thickness, compared to the results obtained with a single, obliquely incident $16\,\rm J$ pulse (solid blue): (a) maximum proton energy and (b) number of protons with $>30\,\rm MeV$ energies. (c) Proton energy distributions from a $40\,\rm nm$ foil using two half-pulses (red), or a single pulse at normal (black) and oblique (blue) incidence. Solid (res. dashed) curves correspond to $l=40\,\rm nm$ (resp. 20~nm). 
}
\label{prot2}
\end{figure}

Compared to that obtained with a single obliquely incident pulse, the optimal target thickness is increased when using two half-pulses, with best results obtained for $l=30-40\,\rm nm$, both regarding the maximum energy [Fig.~\ref{prot2}(a)] and number [Fig.~\ref{prot2}(b)] of accelerated protons. In addition, the proton maximum energy is increased from 119~MeV to 137~MeV. Remarkably, this enhancement goes along with a 3--4-fold rise in the number of high-energy ($>30\,\rm MeV$) protons. When compared with the reference case at normal incidence, the two-pulse setup allows the maximum proton energy to be increased by $>30\%$, and the number of energetic protons to be multiplied by a factor of 4. Although the gain over the single obliquely incident pulse is less impressive, the predicted doubled optimal target thickness could ease the experimental investigation of ion acceleration from relativistically transparent solid foils, given the practical difficulties of manipulating such ultrathin targets. This applies, in particular, to modest-energy laser systems which may not allow achieving RSIT at handleable target thicknesses. Finally, figure~\ref{prot2}(c) shows that the two-pulse scheme leads to substantial improvements in the proton energy spectra over the single-pulse case, both at normal and oblique incidence.

\begin{figure}
\centering
\includegraphics[width=1\textwidth]{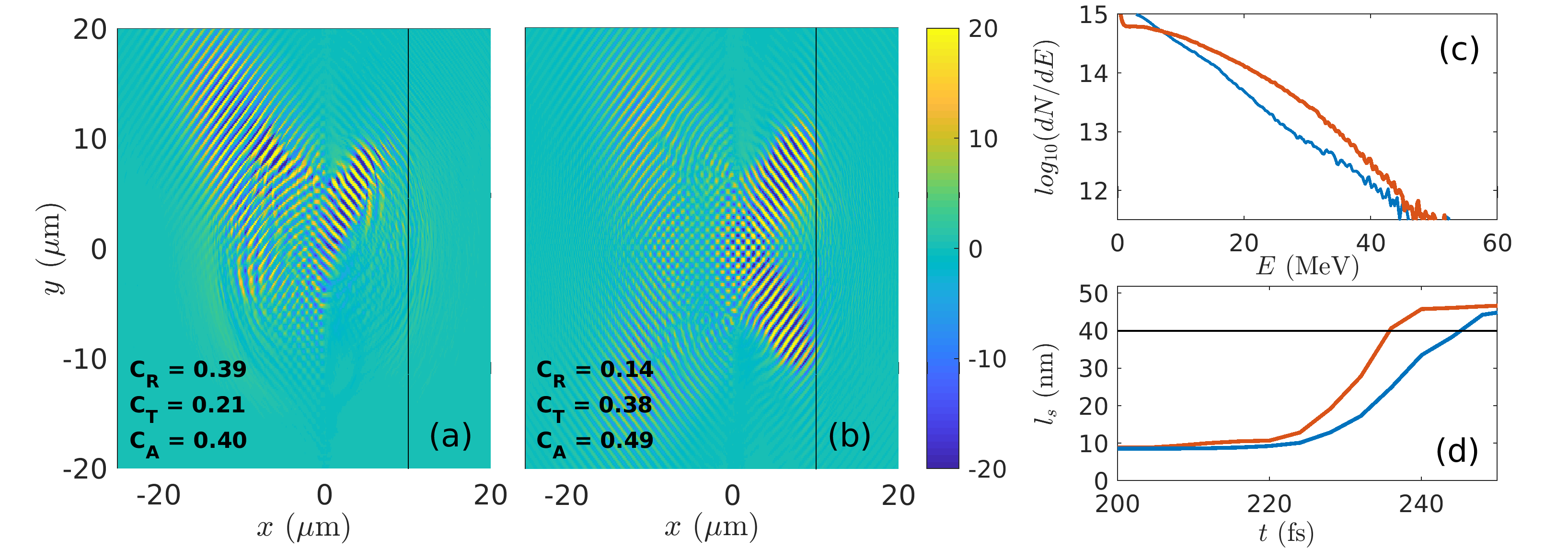}
\caption{(a,b) Transverse electric field $E_y$ (in $\rm V\,m^{-1}$) $35\,\rm fs$ after the on-target laser peak for a single obliquely incident pulse (a) and two half-pulses (b) of same ($16\,\rm J$) total energy. In both cases, the target thickness is set to $40\,\rm nm$, and the measured values of the reflection ($C_R$), transmission ($C_T$) and absorption ($C_A$) coefficients are indicated. The dashed black line delineates the initial position of the target, while the thin grey line serves to highlight that RSIT sets in earlier in (b) than in (a). (c) Electron energy distribution at the peak laser intensity and (d) time evolution of the relativistic skin depth $l_s=\langle\gamma\rangle c/\omega_p$ (with $\langle\gamma\rangle$ the averaged electron Lorentz factor over the laser spot) using a single (blue) or two (red) pulses. The dashed black line indicates the time of the on-target intensity peak. The solid black line indicates the initial target thickness (40~nm).
}
\label{Eyfield1p2p}
\end{figure}

In the two-pulse scheme, the irradiated plasma area is comparable to that associated with a single pulse at oblique incidence. The larger optimum thickness and enhanced ion acceleration then mainly result from the modified electron dynamics in the two-pulse field distribution, 
as has been described earlier in the TNSA regime~\citep{Ferri2019}. For a given target thickness, owing to the stronger electron heating, transparency is then more easily induced using two half-pulses. This is clearly shown in figure~\ref{Eyfield1p2p}, with a transmission coefficient rising from $C_T=0.21$ to $C_T=0.38$ between the single- and two-pulse cases. By examining the laser field distributions in both configurations at various successive times, we checked that RSIT occurs earlier in time using two half-pulses, a result consistent with the $\sim 25\,\%$ increase in the absorption coefficient.

\begin{figure}
\centering
\includegraphics[width=0.45\textwidth]{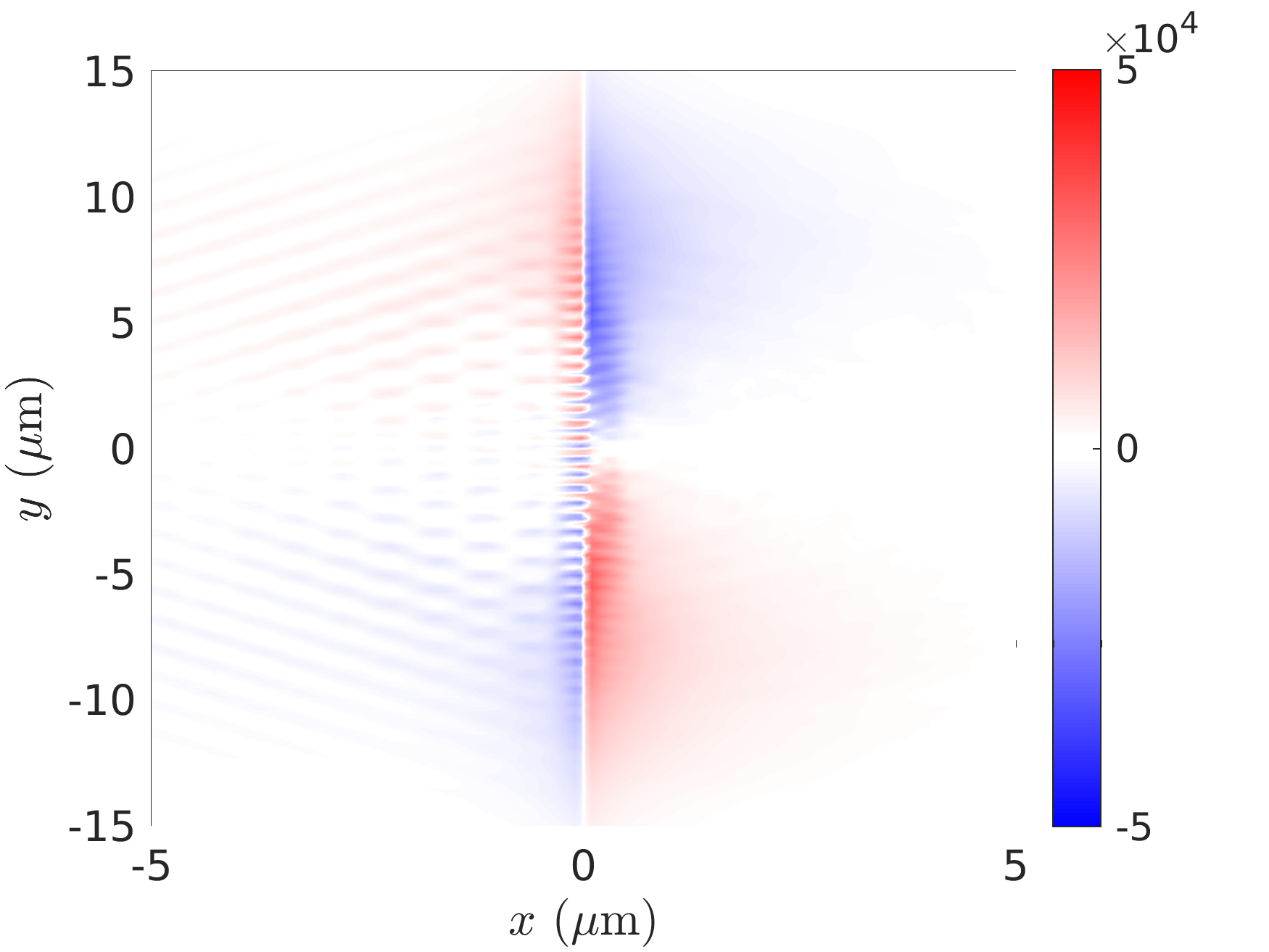}
\caption{Quasistatic $B_z$ field (in T) 10~fs before the on-target intensity peak for two half-pulses interacting with a 40~nm foil. }
\label{Bz220_2p}
\end{figure}

The stronger electron heating [cf Fig. \ref{Eyfield1p2p}(c)] that accounts for the enhanced laser absorption facilitates the induction of transparency. Figure~\ref{Eyfield1p2p}(d) shows the temporal evolution of the relativistic skin depth $l_s=\langle\gamma\rangle c/\omega_p$ during the interaction, with $\omega_p$ the plasma frequency and $\langle\gamma\rangle$ the electron Lorentz factor averaged over the target thickness and around the laser spot. As expected, the use of the two half-pulses yields a faster increase in $l_s$ with the laser intensity.   
We can consider that RSIT sets in when $l_s$ becomes higher than the initial target thickness ($l=40\,\rm nm$). In the two-pulse case, this criterion is fulfilled about $10-15\,\rm fs$ prior to the laser peak time, which incidentally coincides with the onset of RSIT 
with a single pulse. 
We note that this delay between the onset times of RSIT using one or two pulses is consistent with the difference between the transmitted fractions of the laser field in figure~\ref{Eyfield1p2p}.

Furthermore, the symmetry (relative to the $x$ axis) of the two-pulse setup prevents the development of a strong surface magnetostatic field. Figure~\ref{Bz220_2p} shows this field 10~fs before the on-target laser peak for a 40~nm-thick target. As expected from the symmetry of the problem, the $B_z$ field vanishes on axis, and thus cannot counteract the sheath-field-enhancing effect of the stronger laser absorption.

\section{Conclusions}
\indent

Using 2D PIC simulations, we have examined the efficiency of laser-driven proton acceleration in relativistically transparent, ultrathin CH$_2$ foil targets under different short-pulse irradiation conditions. As baseline laser parameters, we have considered a $5\times 10^{20}\,\rm Wcm^{-2}$ maximum intensity and a 38~fs pulse duration. 
First, we have showed that operating at oblique incidence can be beneficial to proton acceleration, owing to enhanced energy absorption into hot electrons.
Yet, the optimum target thickness ($l\simeq 20\,\rm nm$) is then reduced compared to that found at normal incidence ($l \simeq 40\,\rm nm$), as a result of strong ($>10^4\,\rm T$) surface magnetostatic fields self-induced within the ion acceleration region. 
In a second step, we have studied a configuration in which the laser pulse is split into two half-pulses with the same total energy, incoming simultaneously on target but with opposite angles of incidence. This configuration enhances the electron heating while preventing the buildup of deleterious surface $B$-fields within the laser spot. Not only does it allow one to use thicker targets than with a single obliquely incident pulse, but it also yields substantial increases in the proton numbers ($\times 4$) and cutoff energies ($\times 1.3$) over the baseline setup. Our study suggests that the constraints posed by nanometric targets in experiments on RSIT could be somewhat relaxed using a two-pulse scheme. 

\section*{Acknowledgements}
The authors would like to acknowledge fruitful discussions with
I~Thiele and L~Yi. This work was supported by the Knut and Alice Wallenberg Foundation, the Swedish
Research Council, Grant No. 2016-05012, and has received funding from the European Research Council (ERC) under the European Union's Horizon 2020 research and innovation programme under grant agreement No 647121. The simulations were performed
on resources at Chalmers Centre for Computational Science and Engineering (C3SE) provided by the Swedish National Infrastructure for Computing (SNIC).

\bibliographystyle{jpp}

\end{document}